\input{aipcheck}

\documentclass[
    ,final            % use final for the camera ready runs
  ]
  {aipproc}

\layoutstyle{6x9}

\begin{document}

\title{Superdense and normal early-type galaxies at $1<z<2$}

\classification{98.52.Eh; 98.62.Ai; 98.62.Ck; 98.62.Lv; 98.62.Ve}
\keywords      {galaxies: elliptical and lenticular - galaxies: evolution
- galaxies: formation}

\author{P. Saracco}{address={INAF - Osservatorio Astronomico di Brera,
via Brera 28, 20121 Milano, Italy}
}

\author{M. Longhetti}{address={INAF - Osservatorio Astronomico di Brera,
via Brera 28, 20121 Milano, Italy}
}

\author{A. Gargiulo}{address={INAF - Osservatorio Astronomico di Brera,
via Brera 28, 20121 Milano, Italy}
}

\begin{abstract}
 We combined proprietary and archival HST observations to collect 
a sample of 62 early-type galaxies (ETGs) at $0.9<z<2$ with spectroscopic 
confirmation of their redshift and spectral type. The whole sample is 
covered by ACS or NICMOS observations and partially 
by Spitzer and AKARI observations. 
We derived morphological  parameters by fitting their 
HST light profiles and physical parameters by fitting their spectral energy 
distributions.
The study of the size-mass and the size-luminosity relations of these
early-types shows that a large fraction of them ($\sim50\%$) follows 
the local relations.
These 'normal' ETGs are not smaller than local counterparts with comparable 
mass. The remaining half of the sample is composed of compact ETGs with sizes 
(densities) 2.5-3 (15-30) times smaller (higher) than local counterparts and, 
most importantly, than the other normal ETGs at the same redshift and with 
the same stellar mass.  
This suggests that normal and superdense ETGs at $z\sim2$ come from different 
histories of mass assembly.
\end{abstract}

\maketitle

%%%%%%%%%%%%%%%%%%%%%%%%%%%%%%%%%%%%%%%%%%%%
%% MAINMATTER
%%%%%%%%%%%%%%%%%%%%%%%%%%%%%%%%%%%%%%%%%%%%

\section{Introduction}
The presence of early-type galaxies (ETGs) at $z>1$ apparently more compact, 
hence denser, than local ETGs of comparable mass has caught the attention
of many research works in the last 3-4 years. 
Such compact ETGs, often called superdense, are characterized by 
effective radii on average $\sim3$ times smaller than the mean effective 
radius of the local ETGs.
Among the first to point out the smaller radii of high-z
ETGs are \cite{daddi05} who found that 7 ETGs seen in the 
Hubble Ultra Deep Field (HUDF) at $z>1.4$ fall out the local Kormendy 
relation (KR) even taking into account the luminosity evolution.
Subsequent studies of small samples of $z>1$ ETGs based on optical observations
or on seeing limited ground-based observations reached similar results
\cite{dise05}, \cite{truj06} and the first 
deep HST-NICMOS high-resolution (0.075 arcsec/pix) observations in the
near-IR of a sample of $z>1$ ETGs 
confirmed their compactness \cite{long07}. 
The many independent confirmations of the small effective radius of 
high-z ETGs which followed \cite{cima08},  \cite{mcgr08}, 
\cite{buit08}, \cite{damj09} point toward the need of an evolution of the 
effective radius $R_e$ of ETGs  from their redshift to $z=0$.
Recently, \cite{sara09} studying a sample of $\sim30$ ETGs at $1.2<z<1.8$
found that a large fraction of them are actually not more compact than local
ETGs and that those more compact tend to be older.
Here we present the results based on a sample twice the one
studied by \cite{sara09} extending the analysis over a wider range of 
stellar masses and redshift.

\section{The data set}
The sample of ETGs we constructed is composed of 62 ETGs at 
$0.9<z_{spec}<2$ and magnitudes $17<K_{Vega}<20.5$, 28 of which covered 
by HST-NICMOS observations (NICMOS sample hereafter) 
in the F160W filter ($\lambda\sim1.6$ $\mu$m) and 34 by HST-ACS 
observations (ACS sample hereafter).
The NICMOS sample\footnote{Available at 
\url{www.brera.inaf.it/utenti/saracco/32sample.html}} was analysed  
by \cite{sara09} to 
which we refer for a detailed description of the data and of the
various surveys from which they have been extracted.
The ACS sample falls on to the southern field of the  Great Observatories 
Origins Deep Survey  (GOODS-South v2; \cite{giav04}).
The photometry, composed of 14 photometric bands (from UV to mid-IR),  
comes from the GOODS-MUSIC multiwavelength catalog \cite{graz06}
while spectroscopic information come from the ESO-VLT spectroscopic survey 
of the GOODS-South field  (\cite{vanz08} and references therein).

\section{Superdense and normal early-types at $1<z<2$}
We derived the morphological (effective radius) and 
the physical parameters (stellar mass) 
to study the size-mass and the size-luminosity relations for
our sample of ETGs.
The effective radius $r_e$ [arcsec] was derived by fitting a 
S\'ersic profile to the observed galaxy profile 
in the HST NICMOS-F160W and ACS-F850LP images using \texttt{Galfit} software 
(v. 2.0.3). 
Stellar masses $\mathcal{M}_*$  were derived  by fitting the stellar 
population synthesis models of Charlot \& Bruzual (hereafter CB08, 
in preparation) to the observed SED at the spectroscopic redshift $z_{spec}$.
We considered the Chabrier initial mass function (IMF),  
four exponentially declining star formation histories (SFHs) 
$\tau=[0.1, 0.3, 0.4, 06]$ Gyr and metallicity
0.4 $Z_\odot$, $Z_\odot$ and 2 $Z_\odot$ (see \cite{sara09} for a
detailed description of the morphology and SED fitting).
\begin{figure}
  \includegraphics[height=.38\textheight]{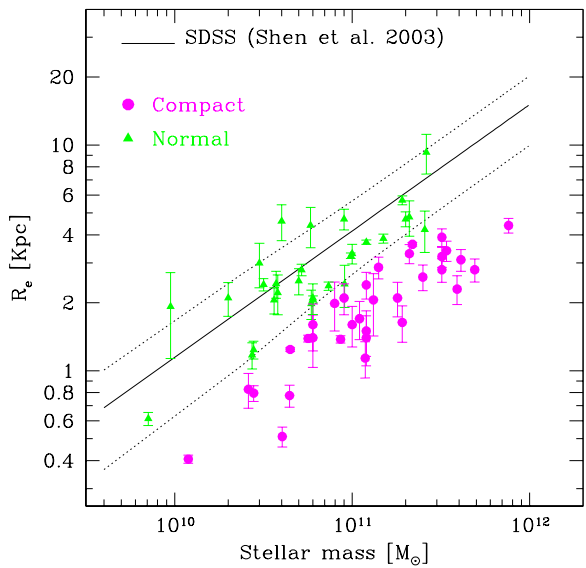}
  \caption{Effective radius R$_e$ versus stellar mass for
our sample of ETGs at $0.9<z<2$. 
The local Size-Mass relation (solid line) found by \cite{shen03} and 
the relevant scatter (dotted lines) are also shown. 
Circles mark compact ETGs having R$_e$ 1$\sigma$ smaller 
than the local relation; triangles mark normal ETGs.}
\end{figure}
In Fig. 1 the size-mass (SM) relation, that is the relation between the 
effective radius R$_e$ [kpc] and the stellar mass $\mathcal{M}_*$ [M$_\odot$] 
of our galaxies
is compared with the SM relation found by \cite{shen03} for the 
local population of ETGs (solid line).
The local SM relation has been scaled by 0.8 to account for
the larger stellar masses provided by the Bruzual and Charlot (2003) 
models with respect to the CB08 models (see \cite{losa09}).
Fig. 1 shows that  $\sim$50\% of the sample (29 out of 62 ETGs, 
filled green triangles)
agrees at $1\sigma$ with the local SM relation, that is it is composed of  ETGs 
having morphological and physical parameters equal to those of local ETGs
(see also \cite{cappe09}, \cite{manci09}).
 The remaining 33 ETGs of the sample (filled purple circles)
diverge  more than one sigma from the local SM relation being them 
more compact than  local ones.
The way in which our sample has been constructed does not allow us to
quantify the real fraction of normal and compact ETGs at $z>1$.
However, it is well known that both imaging and spectroscopic observations 
are biased toward compact galaxies since they have a higher probability to 
be detected thank to the higher S/N ratio. 
Since our sample collects different surveys, it will be biased toward 
compact ETGs too rather than the opposite.
Thus, the actual fraction of normal ETGs at $z>1$ cannot be lower than 
$\sim50$\%, the fraction we observe in our sample and, consequently, the 
population of high-z ETGs is not dominated by galaxies more compact 
than the local ones.

As to the 33 compact ETGs, their effective radius is on average 2.5-3 
times smaller than the mean effective radius of local early-types,
that is their stellar mass density is on average 15-30 times higher.
This offset with respect to the local relation
is constant over the whole mass range spanned by our sample
(from $10^{10}$ M$_\odot$ to $10^{12}$ M$_\odot$), that is no evidence of a 
dependence of the compactness on galaxy mass is found. 
Apart from this, we believe that the most significant result is not 
the possible higher compactness shown by compact ETGs with respect to 
the local population, but rather their higher compactness with respect 
to the normal ETGs at the same redshift.
Indeed, Fig. 1 shows clearly that at $z\sim1.5$ and beyond, when the Universe 
was only 3-4 Gyr old, ETGs fully similar to local ones coexisted with 
other ETGs 15-30 times denser in spite of the same redshift and the same 
stellar mass of all of them.
This result is not dependent on the scaling relation considered.
Indeed, considering the size-luminosity (SL) relation we obtain an 
analogous result. This is shown in Fig. 2 where the SL 
relation, the relation between R$_e$ and the absolute magnitude M$_R$ in 
the R-band of our galaxies, is compared with the local SL relation 
\cite{shen03}.
The absolute magnitude  M$_R$ of our galaxies is the one at the redshift
of the galaxies (left-hand panel). The  offset with respect to the local 
relation reflects the evolution which ETGs undergo from their redshift to 
$z=0$. 
In the right-hand panel of Fig. 2 it is shown how the 62 ETGs of our sample
would be displaced at $z=0$ in the [M$_R$,R$_e$] plane in case of pure
luminosity evolution. 
The agreement between this result and the one derived by the SM relation
(Fig. 1) is remarkable: with very few exceptions, normal ETGs fall within 
one sigma from the local relation and compact ETGs diverge more than one 
sigma from the relation.
This result naturally leads us to conclude that compact and normal 
ETGs result from two different assembly and evolutionary paths
covered at $z>>2$.

\begin{figure}
  \includegraphics[height=.35\textheight]{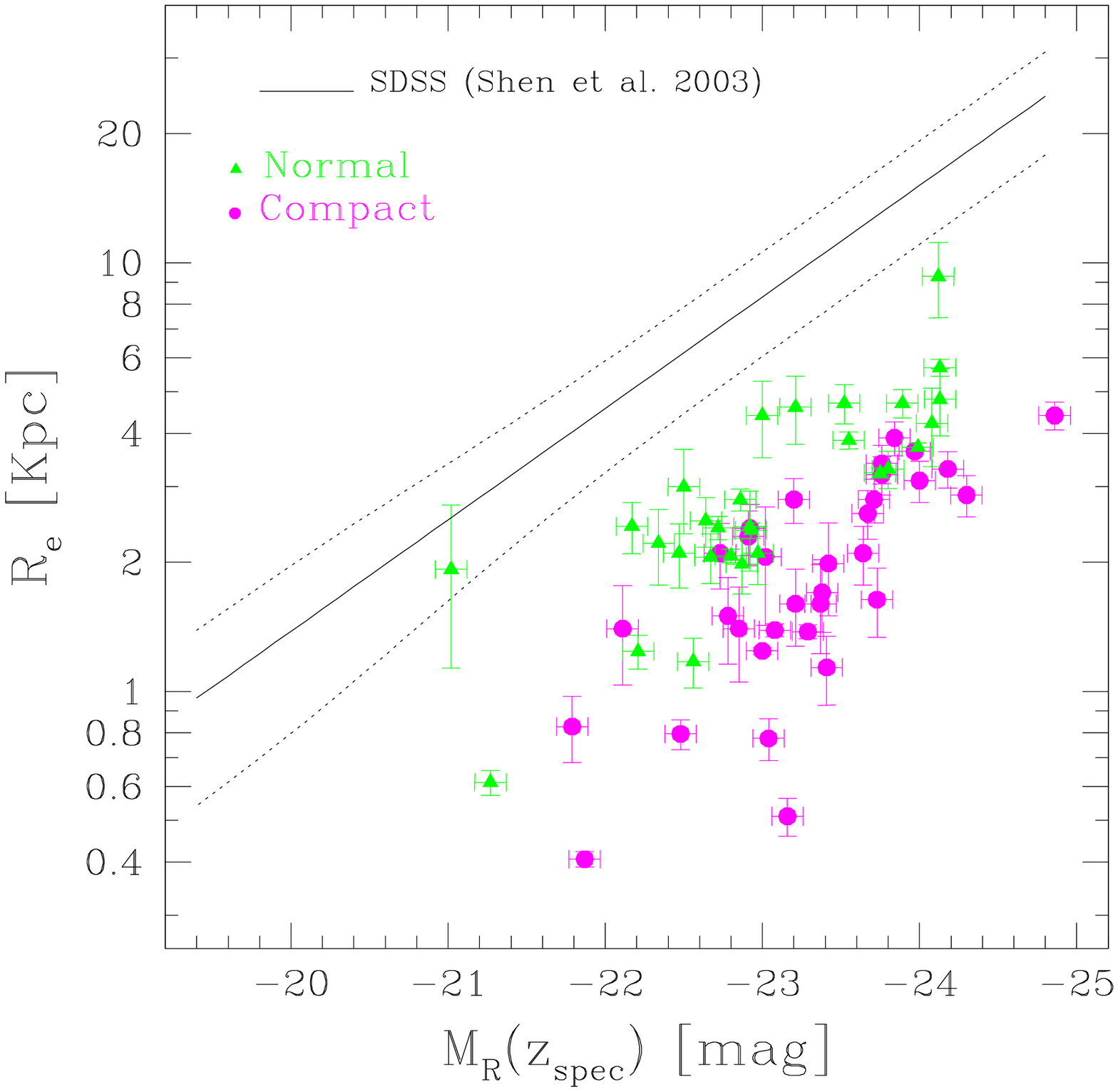}
  \includegraphics[height=.35\textheight]{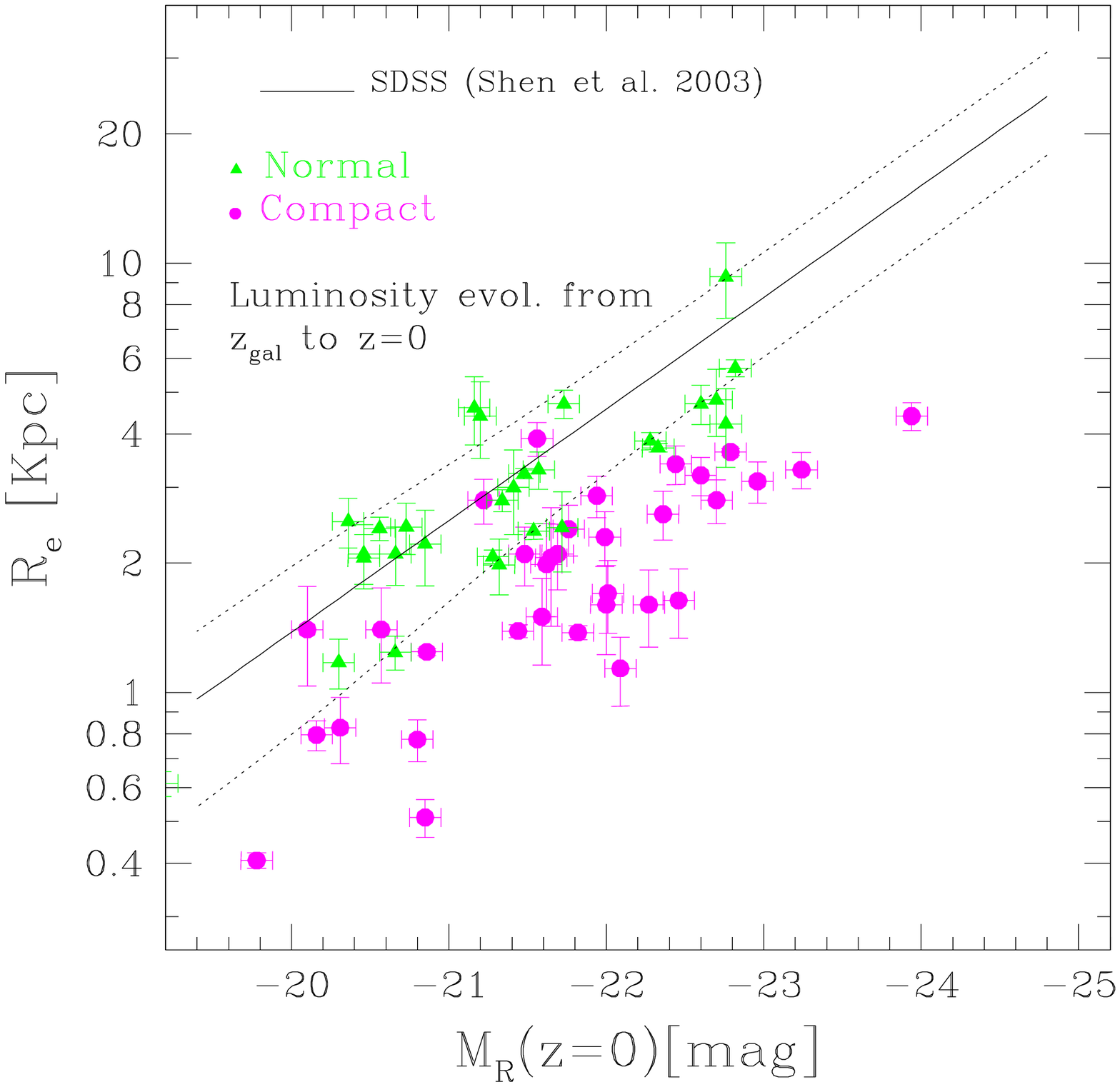}
  \caption{Left-hand panel: SL relation for our sample of ETGs at $0.9<z<2$
  compared with the local relation. Symbols are as in Fig. 1. Right-hand panel:
  SL relation in case of pure luminosity evolution. It is shown how the 62 ETGs
  of our sample would be displaced at $z=0$ in case of pure 
  luminosity evolution}
\end{figure}

%%%%%%%%%%%%%%%%%%%%%%%%%%%%%%%%%%%%%%%%%%%%%%%%
%% BACKMATTER
%%%%%%%%%%%%%%%%%%%%%%%%%%%%%%%%%%%%%%%%%%%%%%%%

\begin{theacknowledgments}
Based on observations made with the NASA/ESA HST, obtained at the Space 
Telescope Science Institute and on observations made with ESO Telescopes at 
Paranal Observatory.
This work has received partial financial support from ASI (contract I/16/07/0).
\end{theacknowledgments}

%%%%%%%%%%%%%%%%%%%%%%%%%%%%%%%%%%%%%%%%%%%%%%%%
%% The bibliography can be prepared using the BibTeX program or
%% manually.
%%
%% The code below assumes that BibTeX is used.  If the bibliography is
%% produced without BibTeX comment out the following lines and see the
%% aipguide.pdf for further information.
%%
%% For your convenience a manually coded example is appended
%% after the \end{document}
%%%%%%%%%%%%%%%%%%%%%%%%%%%%%%%%%%%%%%%%%%%%%%%%

%%%%%%%%%%%%%%%%%%%%%%%%%%%%%%%%%%%%%%%%%%%%%%%%
%% You may have to change the BibTeX style below, depending on your
%% setup or preferences.
%%
%%
%% For The AIP proceedings layouts use either
%%%%%%%%%%%%%%%%%%%%%%%%%%%%%%%%%%%%%%%%%%%%

\bibliographystyle{aipproc}   % if natbib is available

\begin{thebibliography}{9}
\bibitem{daddi05}
Daddi E., Renzini A., Pirzkal N., et al.,\emph{ApJ}, \textbf{626}, 680
(2005)
%\bibitem{cassa05}
%Cassata P., et al., \emph{MNRAS}, \textbf{357}, 903, (2005)

\bibitem{dise05}
di Serego Alighieri S., et al., \emph{A\&A}, \textbf{442}, 125 (2005)

\bibitem{truj06}
Trujillo I., Feulner G., Goranova Y., et al., \emph{MNRAS}, \textbf{373}, L36
(2006)

\bibitem{long07}
Longhetti M., Saracco M., Severgnini P., et al., \emph{MNRAS}, \textbf{374}, 614
(2007) 


\bibitem{cima08} 
Cimatti A., Cassata P., Pozzetti L., et al., \emph{A\&A}, \textbf{482},
21 (2008)

\bibitem{mcgr08} 
McGrath E. J., Stockton A., Canalizo G., Iye M., Maihara T.,
\emph{ApJ}, \textbf{682}, 303 (2008) 


\bibitem{buit08} 
Buitrago F., Trujillo I., Conselice C. J., M., et al., \emph{ApJL}, \textbf{687},
61 (2008)

\bibitem{damj09} 
Damjanov I., McCarthy P. J., Abraham R. G., et al., \emph{ApJ}, \textbf{695},
101 (2009)

\bibitem{sara09}
Saracco P., Longhetti M., Andreon S., \emph{MNRAS}, \textbf{392}, 718 (2009)

\bibitem{giav04}
Giavalisco, M., Dickinson, M., Ferguson, H. C., et al., \emph{ApJL}, 
\textbf{600}, 103 (2004)

\bibitem{graz06}
Grazian, A., Fontana, A., de Santis, C., et al., \emph{A\&A}, \textbf{449},
951 (2006)

\bibitem{vanz08}
Vanzella E., Giavalisco M., Dickinson M., et al., \emph{A\&A}, \textbf{478},
83 (2008)


\bibitem{losa09}
Longhetti M., Saracco P., \emph{MNRAS}, \textbf{394}, 774 (2009)

\bibitem{shen03}
Shen S., et al., \emph{MNRAS}, \textbf{343}, 978 (2003)

\bibitem{cappe09}
Cappellari M., et al., \emph{ApJ}, \textbf{704}, L34 (2009)
\bibitem{manci09}
Mancini C., Daddi E., Renzini A., et al., \emph{MNRAS}, in press,  {arXiv:0909.3088}
(2009)
\end{thebibliography}

\end{document}